\documentclass[final,1p,times,twocolumn]{elsarticle}

\usepackage{graphicx}

\journal{Nuclear Physics A}

\begin{document} 
\newcommand{\sigmaTotal}{\sigma_{tot}}
\renewcommand{\Re}{\mathrm{Re\,}}
\renewcommand{\Im}{\mathrm{Im\,}}

\begin{frontmatter}
\title{Proton periphery activated by multiparticle dynamics} 

\author[lpi]{I.M. Dremin}
 \ead{dremin@lpi.ru}
\author[lpi,itep]{V.A. Nechitailo}%

\address[lpi]{%
 P.N.~Lebedev Physical Institute, Moscow 119991, Russia
}%

\address[itep]{Institute of Theoretical and Experimental Physics, Moscow, Russia}

\begin{abstract}
It is shown that protons become more active at the periphery with increase
of their collision energy. By computing the impact parameter distribution of
the proton-proton overlap function at LHC energies and comparing it with
ISR (and S$p\bar p$S for $p\bar p$) data, we conclude that the peripheral
region of protons plays an increasing role in the rise of total cross sections
through multiparticle dynamics. The size of the proton as well as its blackness 
increase with energy. The protons become more black both in the central region 
and, especially, at the periphery. This effect can be related to the ridge
phenomenon and to the inelastic diffraction processes at LHC energies.
\end{abstract}
\end{frontmatter}

It is well known that parton (quark, gluon) densities and the share of
low-$x$ partons rise with increasing energies of colliding hadrons. Less
attention has been paid to the analysis of the spatial distribution of the
parton content inside them and its evolution with energy. This can be done by
studying the structure of the overlap function in the unitarity condition
for the elastic scattering amplitude in the impact parameter representation
at different energies of colliding protons. The very first analyses 
\cite{hv74, htk74, as80} have lead to extremely interesting conclusions about 
the increasing peripherality of protons within the rather narrow interval of 
ISR energies. Later, this was confirmed and strengthened at somewhat higher 
energies by the S$p\bar p$S data \cite{4}. It was shown that while the increase 
of the overlap function at the proton periphery is quite modest in the ISR energy
range (about 4$\%$), it becomes much stronger (about 12$\%$) if S$p\bar p$S
energies are included. However, no sizeable change of the proton blackness was
noticed at small distances in this energy interval. The similar effect at HERA
energies was discussed in \cite{msmu} for the vector meson production process
in the framework of the dipole-proton scattering model.

That is why we attempt to learn if peripheral regions of protons become even 
more active at LHC energies and the central region is activated as well. The 
striking, but not at all unexpected, result is that this increase persists 
and extends to smaller impact parameters now. It amounts to about 40$\%$ of 
edge corrections at distances about 1 fm. The main parton content in the 
overall region of inelastic collisions remains relatively constant and below 
the unitarity bound in the central region of impact parameters less than about 
0.5 fm but also indicates some increase of opacity compared to lower energies.

We proceed by using the approach adopted in \cite{as80}. First of all, the TOTEM
data on the differential cross section of elastic pp-scattering at 7~TeV 
\cite{totem} are fitted by the formula (\ref{ampl}) for the elastic scattering
amplitude $f(s,t)$ (which depends on the center-of-mass energy $\sqrt s$ and
the transferred momentum $\sqrt {\vert t\vert }$) proposed in \cite{nagy}: 
\begin{equation}
f(s,t)=i\alpha[A_1\exp (\frac{1}{2}b_1\alpha t)+A_2\exp ( \frac{1}{2}b_2\alpha t )]-
iA_3\exp (\frac{1}{2}b_3 t),
\label{ampl}
\end{equation}
where $\alpha (s)$ is complex and is given by
\begin{equation}
\alpha (s)=[\sigmaTotal(s)/\sigmaTotal(23.5\, \mathrm{GeV})](1-i\rho_0(s)).
\label{alph}
\end{equation}
Even though this parametrization has no theoretical foundation, it provides
in a wide energy interval good phenomenological fits of differential cross 
sections\footnote{There are many others (albeit with larger number of 
adjustable and hidden parameters) reviewed in \cite{ufn} and recently 
published \cite{bmsw, flmn, kfk1, kfk2}.} defined as
\begin{equation}
\frac {d\sigma }{dt}=\vert f(s,t)\vert ^2.
\label{dsigma}
\end{equation}
The normalization at the optical point is
\begin{equation}  
\sigmaTotal(s)=\sqrt {16\pi }\Im f(s,0).
\label{norm}
\end{equation}
We shall also use the ratio of the real to imaginary parts of the amplitude
\begin{equation}
\rho (s,t)=\frac {\Re f(s,t)}{\Im f(s,t)}.
\label{rho0}
\end{equation}
The following values of the parameters have been fixed by the fit to the 
experimental points of the differential cross section at the energy 7~TeV
in the range $0.0052<\vert t\vert <2.44\, \mathrm{GeV}^2$:
\begin{eqnarray}
A_1^2 = 55.09 \, \mathrm{mb/GeV}^2, & A_2^2 = 3.46 \, \mathrm{mb/GeV}^2, &
A_3^2 = 1.47 \, \mathrm{mb/GeV}^2, \nonumber\\
b_1 = 8.31\, \mathrm{GeV}^{-2}, &  b_2 = 4.58\, \mathrm{GeV}^{-2}, & b_3 = 4.70\, \mathrm{GeV}^{-2}.
\label{para}
\end{eqnarray}
We have also used $\sigmaTotal(7\,\mathrm{TeV})=98.58 \, \mathrm{mb}$  and 
$\rho _0(7 \, \mathrm{TeV})=0.14$ which leads to $\rho(s,0)=0.148$. 
The parameter $\rho _0$ defines $\rho(s,0)$ but not coincides with it.
The normalization of $\vert f(s,t)\vert ^2$ in 
mb/GeV$^{2}$ allows direct comparison with \cite{as80}. Note that the exponentials in first two terms are
determined by the product of $b_i$ and $\alpha (s)$.

\begin{figure}
\includegraphics[width=\textwidth]{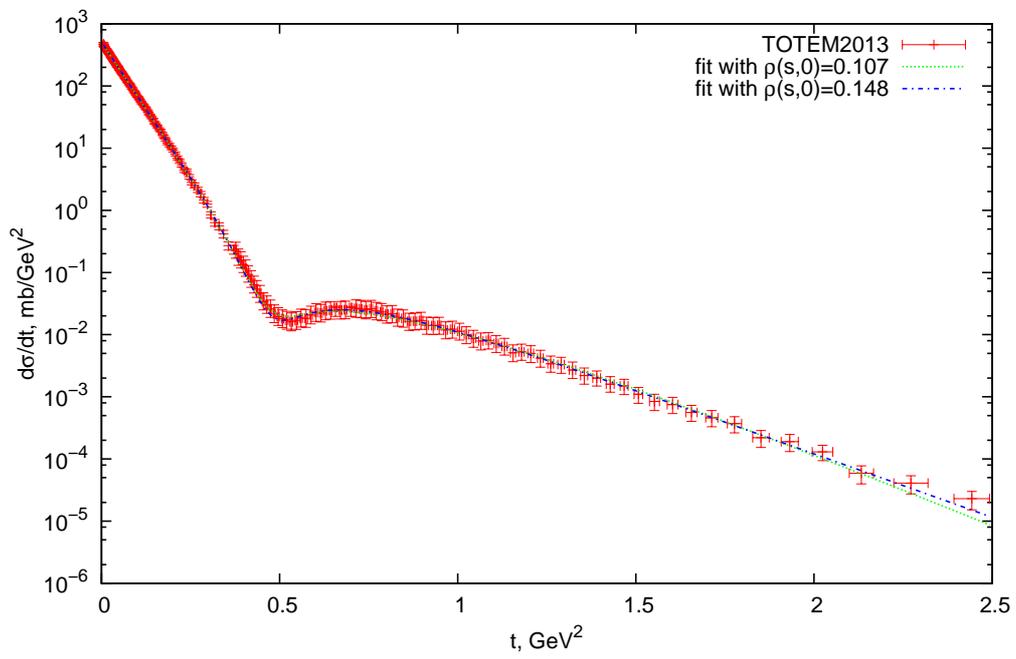}
\caption{Fit of the TOTEM data -- dotted and dash-dotted curves. Dotted curve is calculated with parameter
$\rho(s,0)=0.107$ and dash-dotted curve with $\rho(s,0)=0.148$}
 \label{TOTEM_Amaldi_fit}
\end{figure}

The good quality of the fit with these values is seen in Fig.~\ref{TOTEM_Amaldi_fit}. 
A complete error analysis would lie beyond the scope of this paper. We have 
just estimated that uncertainties are small by varying the measured quantities 
inside the $\pm 1\sigma$ error bars.
Thus, we consider the formula (\ref{ampl}) as an empirical insert suitable for 
the numerical fit of experimental data and use it in further calculations.
The fit is almost insensitive to the parameter 
$\rho _0$ except of the dip region. With some adjustment of parameters 
$A_i, b_i$, the fit is also satisfactory if one adopts the value of 
$\rho(s,0)=0.107$ favored by the recent results of the TOTEM collaboration 
\cite{ost}.

\begin{figure}
\includegraphics[width=\textwidth]{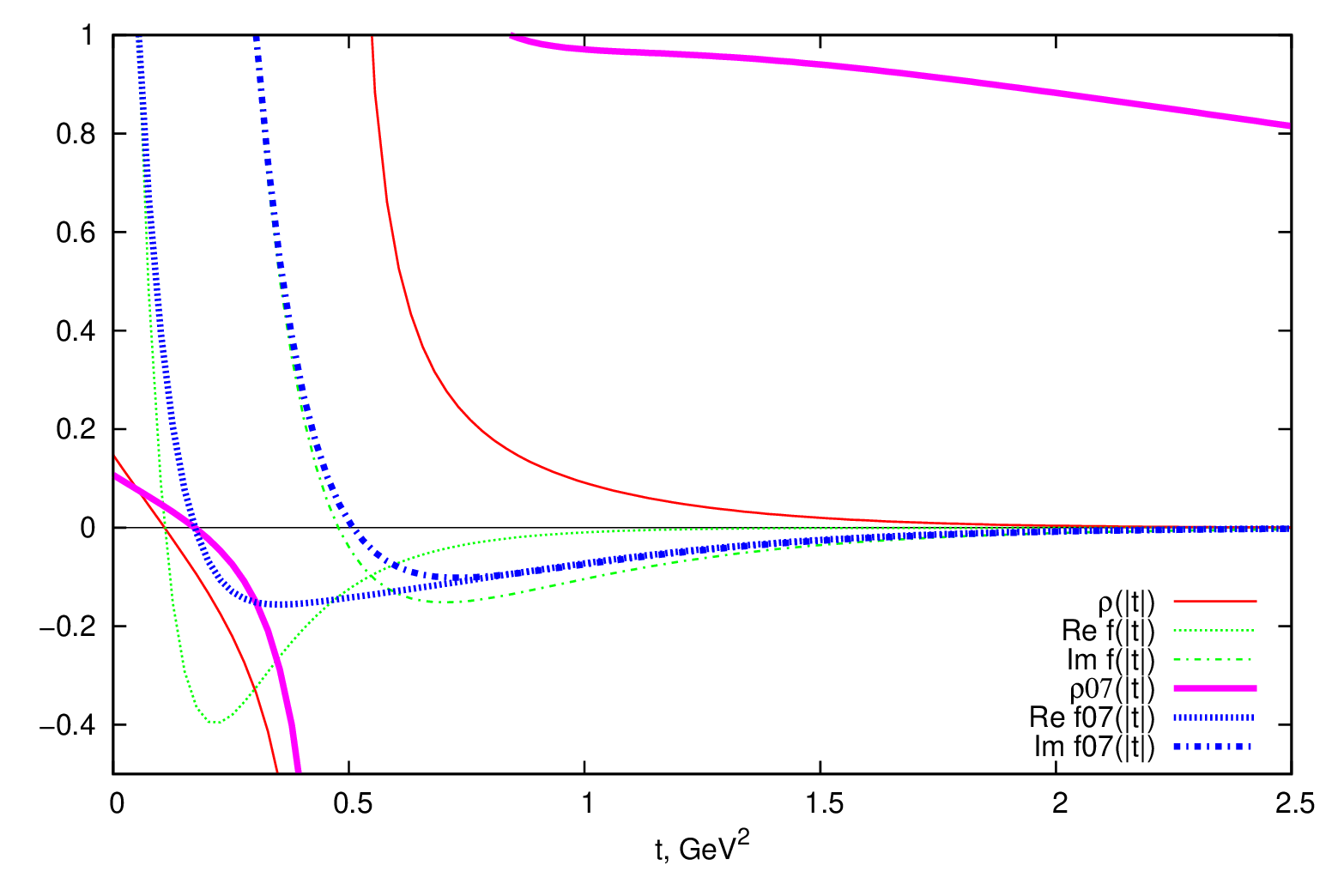}
\caption{Real (dotted curve) and imaginary (dash-dotted curve) parts of the amplitude and their ratio (solid curve).
Bold lines are for $\rho(s,0)=0.107$.}
 \label{TOTEM_rho_ReIm}
\end{figure}

One is able to compute from (\ref{ampl}) the real and imaginary parts of the 
amplitude at any value of the transferred momentum $t$ using these parameters.
They are shown in Fig.~\ref{TOTEM_rho_ReIm}. Each of them has a single zero. The evolution of their 
ratio $\rho (s,t)$ with the transferred momentum $t$ is mainly prescribed by 
the last term in this formula with the negative sign in front of it and small 
exponential $b_3$. The ratio
$\rho (s,t)$ for $\sqrt s$=7~TeV is also shown in Fig.~\ref{TOTEM_rho_ReIm}. It clearly 
demonstrates that the real part is small and can be neglected everywhere 
except near the point where the imaginary part is equal to zero.

The new feature seen in Fig.~\ref{TOTEM_rho_ReIm} is the difference in values of $\rho (s,t)$
in the Orear region for the two choices of $\rho(s,0)$. It becomes of the
order of 1 for $\rho(s,0)=0.107$. Note that in both cases it strongly differs
from its average value about -2 required by the fit according to the solution 
of the unitarity equation \cite{dnec}. The solution \cite{adre} predicts 
the exponential of $\sqrt {\vert t\vert }$-behavior of the amplitude. It 
poses an interesting problem
which has not yet been resolved. 

This ratio can change the sign if more zeros
of either imaginary or real parts of the amplitude appear. For example, this
happens in the model of \cite{kfk1}. It incorporates phenomenologically
the Orear type behavior of the amplitude at larger thansferred momenta 
(albeit in a way somewhat different from \cite{adre}) and predicts two zeros 
of the real part. Then, the ratio reaches large negative values at high 
transferred momenta. It can solve the above puzzle. The somewhat smaller 
absolute values at large $\vert t\vert $
are also predicted in the model using the inverse of polynomials \cite{bmsw}
as described in \cite{kfk1}. The analysis of \cite{kfk1} clearly shows that the 
somewhat modified description of the amplitude at larger transferred momenta 
can change our conclusions about the behavior of this ratio there.

To reveal the space structure of proton interactions, the amplitude $f(s,t)$
must be rewritten in the impact parameter space in place of the momentum 
space. By applying the Fourier-Bessel transformation, we define the 
dimensionless profile function
\begin{equation}
i\Gamma(s,b)=\frac{1}{\sqrt{\pi }}\int _0^{\infty}dq q f(s,t) J_0(qb).
\label{gamm}
\end{equation}
Here, the variable $b$, called the impact parameter, describes the vector
joining the centers of colliding protons at the moment of their collision,
$q=\sqrt{-t}$, and $J_0$ is the Bessel function of zero order.

The amplitude $f(s,t)$ must satisfy the unitarity condition. For $t=0$, it
states that the total cross section is a sum of cross sections of elastic and
inelastic processes. If written in the
impact parameter representation (\ref{gamm}) it looks like
\begin{equation}
2\Re\Gamma (s,b)=\vert \Gamma (s,b)\vert ^2+G(s,b),
\label{unit}
\end{equation}
where $G$ is called the overlap function in the impact parameter space.

The smallness of the real part of $f(t)$ corresponding to small
$\Im\Gamma(s,b)$ implies that one can compute $G$ approximately as
\begin{equation}
G(s,b)\approx 2\Re\Gamma(s,b)-(\Re\Gamma(s,b))^2.
\label{over}
\end{equation}
The physics meaning of these relations is very simple. The overlap function $G$
describes the kinematical overlap of two cones filled in by the inelastically 
produced secondary particles in the momentum space expressed in terms of the 
proton structure
at a given impact parameter $b$. In other words, it corresponds to the particle 
distribution $d^2\sigma /d \mathbf{b}$ in the impact parameter space. One may treat it
as a parton distribution if one-to-one correspondence of particles and partons 
is assumed.

\begin{figure}
\includegraphics[width=\textwidth]{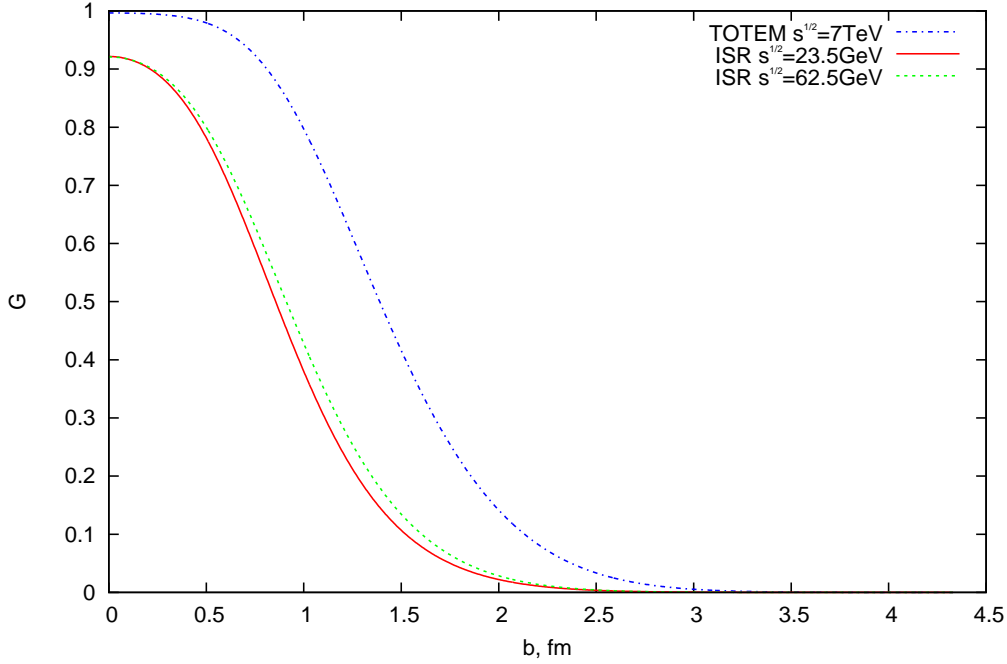}
\caption{The overlap functions at 23.5~GeV (solid curve), 62.5~GeV (dotted curve) and 7~TeV (dash-dotted curve)}
\label{overlap_function}
\end{figure}

The overlap function at 7~TeV has a form shown in Fig.~\ref{overlap_function} by the upper 
dash-dotted curve. Its dependence on $\rho(s,0)$ is so weak that can be neglected.
However, as we see, it strongly differs from the corresponding function at ISR 
energy 23.5~GeV (shown by the lower solid curve), especially at the very edge 
of the distribution. The overlap function at 7~TeV declines steeply but there is no sharp 
cutoff at large impact parameters. At $b=0$, it approaches
the unitarity limit corresponding to the complete blackness. This is a clear 
manifestation of the parton saturation effect.

The difference between the two functions $\Delta G(b)=G(s_1,b)-G(s_2,b) \quad (\sqrt{s_1}=7\, \mathrm{TeV}$, 
$\sqrt {s_2}=23.5\, \mathrm{GeV}$)
results because of
the increase of the elastic cross section and shrinkage of the diffraction cone
with energy. In other words, it demonstrates the increase of the opacity since
the ratio $\sigma _{el}/\sigmaTotal$ increases also, and it is proportional to
the opacity.

\begin{figure}
\includegraphics[width=\textwidth]{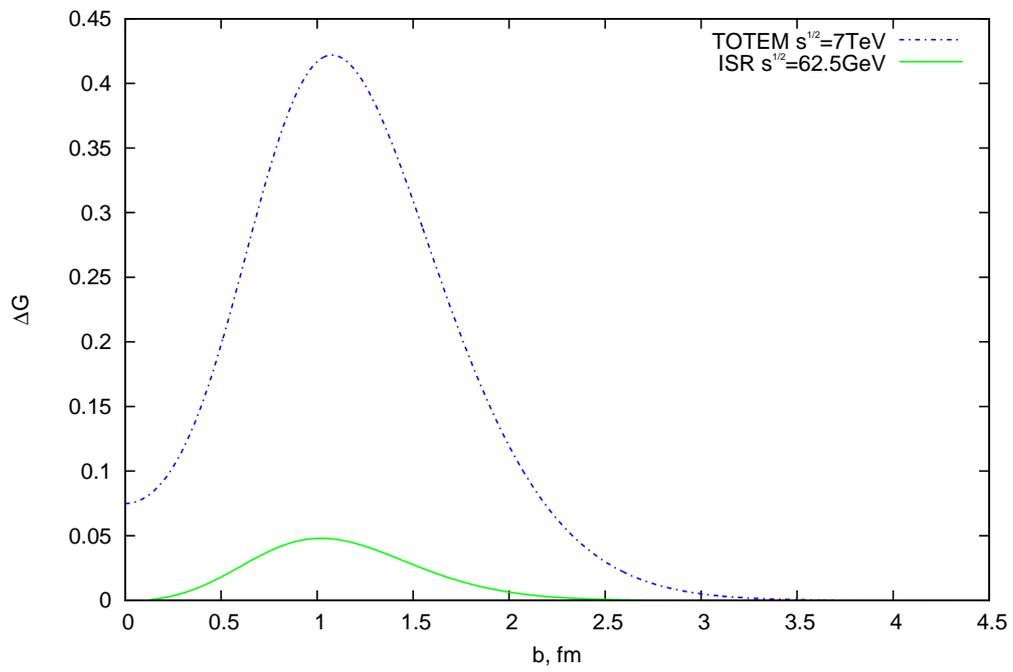}
\caption{The difference between the overlap functions.  Dash-dotted curve is for 7~TeV and 23.5~GeV energies,
solid curve is for 62.5~GeV and 23.5~GeV energies}
\label{overlap_diff}
\end{figure}

In Fig.~\ref{overlap_diff}, we demonstrate the difference between these two distributions (the 
upper curve). It is mainly concentrated at the periphery of the proton at the 
distance about 1~fm. This feature is stable against the variations of $\rho(s,0)$.
It shows that, at higher energies, the peripheral region 
becomes more populated by partons, and they play more active role in particle 
production. 

It is tempting to ascribe the peripheral nature of this effect to two features
of inelastic processes observed already at LHC. First, the collisions with
impact parameters about 1 fm lead to the almond-shaped overlap region. Therefore,
due to increase of the parton density they become responsible for the 
ridge-effect visible in high multiplicity pp-processes at LHC but not observed 
at lower energies. Second, the more peripheral collisions with larger impact 
parameters would lead to strong increase of the cross section of the inelastic 
diffraction with large masses and high multiplicities which can hardly be 
separated by the gap criteria from the minimum bias events. This is especially 
interesting because the cross section of the low-mass diffraction is rather 
small at 7~TeV \cite{atot} and surprisingly close to its values at ISR energies.
The stronger absorption in the peripheral region at 7~TeV results in the 
suppression of the low-mass inelastic diffraction processes. It looks as if it 
is necessary to include the states of the continuum spectrum beside the discrete
bare eigenstates in the traditional approach \cite{gwal}.

We note that the Regge-type models of inelastic
diffraction \cite{kaid} do not agree with the observed decrease below 1 at 7~TeV 
of the parameter $Z=4\pi B/\sigmaTotal$ studied in \cite{ufn} because they predict 
that this parameter is equal to $1+\sigma_{in}^D/\sigma_{el}$ greater than 1
(where $\sigma_{in}^D$ is the cross section of the low-mass inelastic 
diffraction). It has been pointed out in \cite{adre} that this parameter
defines the slope of $d\sigma/dt$ beyond the diffraction cone in the Orear 
region. Its further decrease with energy would result \cite{drem} in first
signatures of the approach to the black disk limit to appear just in there.

Another new feature, seen in Fig.~\ref{overlap_diff}, is the quite large (about 0.08)
value of $\Delta G(b)$ at small impact parameters that reveals a stronger
blackness of the disk at higher energies and is related to the rise of the cross 
section of the main bulk of inelastic processes. No signs of this effect have 
been found at lower energies. These observations are tightly related to the 
visible violation of geometric scaling even in the diffraction cone at LHC 
energies \cite{drne} because of the dual correspondence of the transferred 
momenta $t$ and the impact parameters $b$. Probably, they correspond also
to disagreement between experimental data at 7~TeV and predictions of 
Monte Carlo models seen in high multiplicity events \cite{10022}. 

In the same Figure, the lower curve corresponds to the similar (albeit much
smaller!) difference $\Delta G(b)$ within the quite narrow ISR energy interval 
(23.5 - 62.5~GeV). It was stressed in \cite{as80} that this difference is
negligibly small at low impact parameters while showing some statistically 
significant excess about 4$\%$ at the periphery\footnote{This curve differs
slightly from that of \cite{as80} because we did not use the interpolation 
procedure adopted there but subtracted directly two overlap functions at
62.5~GeV and 23.5~GeV.}.

We stress that the formula (\ref{ampl}) has been applied just for analytic
approximation of experimental data with the smallest number of adjustable 
parameters. Its extrapolation to the yet unmeasured
regions of $d\sigma /dt$ can hardly change the main conclusion qualitatively. The range of
$\vert t\vert $ below 0.005~GeV$^2$ ($p_T<70\,\mathrm{MeV}$; large $b$) is well fitted by
the normalization to the optical point. In the range of $\vert t\vert > 2.5\, \mathrm{GeV}^2$,
the values of $d\sigma /dt$ become extremely small to 
seriously influence the qualitative conclusions about the 
energy evolution of the spatial picture described above. 
However, we should point out that the behavior at larger transferred momenta 
is important for understanding the relative role of real and imaginary parts
and can somewhat change the quantitative estimates of $\Delta G$.
Also, the slight variations of the overlap function at low $b$ are crucial 
for theoretical modelling.

Comparison of the curves in Fig.~\ref{overlap_diff} leads to the conclusion that the
size of the proton as well as its blackness increase with energy. The protons
become more black both in the central region, where they almost reach the 
saturation of the unitarity condition, and, especially, at the periphery,
where the parton density strongly increases.
Multiparticle dynamics is in charge of these effects. 

\section*{Acknowledgments}

This work was supported by the RFBR grant 12-02-91504-CERN-a and
the RAS-CERN program.                                   

\bibliographystyle{model1b-num-names}
\bibliography{peri}

\end{document}